\documentclass[10pt,aps,prd,fleqn,superscriptaddress,twocolumn,nofootinbib,preprintnumbers]{revtex4-1}
\pdfoutput=1
\usepackage{hyperref,amsmath,amssymb,nicematrix,graphicx,nicefrac}
\hypersetup{
  pdfauthor={Enrico Bothmann,John Campbell,Stefan Hoeche,Max Knobbe},
  pdftitle={Algorithms for numerically stable scattering amplitudes}
}

\begin{document}
\preprint{FERMILAB-PUB-24-0272-T, MCNET-24-10}
\title{Algorithms for numerically stable scattering amplitudes}
\author{Enrico Bothmann}
\affiliation{Institut f\"ur Theoretische Physik, Georg-August-Universit\"at G\"ottingen, 37077 G\"ottingen, Germany}
\author{John~M.~Campbell}
\affiliation{Fermi National Accelerator Laboratory, Batavia, IL, 60510, USA}
\author{Stefan~H{\"o}che}
\affiliation{Fermi National Accelerator Laboratory, Batavia, IL, 60510, USA}
\author{Max Knobbe}
\affiliation{Institut f\"ur Theoretische Physik, Georg-August-Universit\"at G\"ottingen, 37077 G\"ottingen, Germany}

\begin{abstract}
The numerically stable evaluation of scattering matrix elements near the infrared limit of
gauge theories is of great importance for the success of collider physics experiments.
We present a novel algorithm that utilizes double-precision arithmetic and reaches
higher precision than a naive quadruple-precision implementation at smaller computational cost.
The method is based on physics-driven modifications to propagators, vertices and 
external polarizations.
\end{abstract}
\maketitle

During the past decades, particle physics experiments have explored the structure of matter and
its interactions at an unprecedented pace and precision. We owe much of this scientific progress to 
the development of particle colliders~\cite{EuropeanStrategyforParticlePhysicsPreparatoryGroup:2019qin,
  *Butler:2023glv}. The physics program at collider experiments depends critically
on theoretical predictions based on the Standard Model of particle physics, which are typically
obtained using perturbation theory~\cite{Narain:2022qud,*Craig:2022cef}.

At higher orders in the perturbative expansion, scattering matrix elements develop 
infrared singularities that arise from long-distance interactions between degenerate
asymptotic states~\cite{Bloch:1937pw,*Yennie:1961ad,*Kinoshita:1962ur,*Lee:1964is}.
These infrared singularities lead to enormous complications in numerical calculations
that rely on Monte-Carlo methods for the integration over phase space. They have become
particularly problematic in the context of semi-automated computations at higher order in QCD,
which are needed to realize the full physics potential of the
Large Hadron Collider~\cite{Heinrich:2020ybq,*FebresCordero:2022psq,Huss:2022ful}.
At the next-to-leading order, calculations can be performed with the help of generic 
infrared subtraction methods~\cite{Frixione:1995ms,*Catani:1996vz,*Catani:2002hc}.
A similar automation at the next-to-next-to leading order in QCD seems
within reach~\cite{Huss:2022ful}.
All existing methods do, however, rely on the cancellation between real-emission 
matrix elements, and their corresponding approximations in the soft and collinear
limits~\cite{Mueller:1981ex,*Ermolaev:1981cm,*Dokshitzer:1982fh,*Dokshitzer:1982xr,*Bassetto:1982ma,
  *Bassetto:1984ik,Dokshitzer:1977sg,*Gribov:1972ri,*Lipatov:1974qm,*Altarelli:1977zs}.
Approaching the limit, the finite remainder can only be computed reliably if both
the matrix elements and their approximations can be evaluated at high
numerical accuracy {\it individually}.

It is often assumed that numerical stability issues in the computation of matrix elements
are best addressed with the help of multiple precision arithmetic. In this letter,
we will show that this is neither necessary nor sufficient. We propose a novel method
that is computationally more efficient and also offers the opportunity to probe the deep
infrared region through a suitable reformulation of the relevant components of
tree-level matrix elements. While the details of a numerical implementation can be
quite involved, the basic idea behind the method is simple: We express every part of the
amplitude in terms of large and small components of four momenta. Quantities that
become parametrically small in the infrared limits appear in most cases in the form of
ratios of the small over the large momentum components, and such ratios can always be
evaluated reliably. This is in stark contrast to a naive multiple-precision computation, 
where small quantities typically emerge as the difference between components of four momenta
that are individually large. At no point in the calculation does our new method involve
any approximations. All expressions remain exact to the given order in perturbation theory.

\section{Stability problems and solutions}
\label{sec:expansions}
Numerical imprecisions in tree-level matrix elements arise from multiple sources.
The one that is most easily understood is the denominator of propagators in momentum space.
For massless particles, it contains terms of the form $1-\cos\theta$, which originate in the
retarded Greens functions of the theory. In this context, $\theta$ is the opening angle between
the momentum of the particle emitting a gauge field quantum, and the momentum of the emission.
Whether using Feynman graphs, or a recursive algorithm such as the Berends-Giele method~\cite{
  BERENDS1988759,*Berends:1988yn,*Berends:1990ax}, we need to find a numerically stable
  algorithm to compute such terms when $\cos\theta\approx 1$. This is relatively straightforward.
The numerators of Feynman diagrams, which arise from the spin-dependence of the interactions,
are harder to control. There are often intricate cancellations among different graphs,
which are difficult to achieve numerically. To solve this problem, we will propose
convenient expressions for the vertex factors and external wave functions, and make use
of a physical gauge that suppresses interferences in the collinear 
limits~\cite{Amati:1978wx,*Amati:1978by,*Ellis:1978sf,*Ellis:1978ty,
  *Kalinowski:1980ju,*Kalinowski:1980wea,*Gunion:1985pg,*Catani:1999ss}.

\subsection{Scalar products and light-cone momenta}
\label{sec:dot_products}
We begin by discussing a numerically stable algorithm to compute the invariant
mass of a momentum vector $q=p+k$, which is given by $q^2=p^2+k^2+2pk$. The key idea is
to represent each four-momentum in the numerical calculation as a five-vector, with the
fifth component being the virtuality. If this virtuality is known to high precision, the
task of finding $q^2$ reduces to a computation of the scalar product, $2pk$. In anticipation
of the expressions needed to evaluate vertex factors involving vector boson currents, we will
discuss the most general case of complex four-momenta. 
The scalar product can be rewritten as
\begin{equation}\label{eq:dot_product}
  \begin{split}
    2pk
    =&\;2p_0k_0\bigg[\,1-\frac{\bar{p}\,\bar{k}}{p_0k_0}\,\bigg]
    +2\,\bar{p}\,\bar{k}\,\Big(1-\vec{v}_p\vec{v}_k\Big)\;,
  \end{split}
\end{equation}
where
\begin{equation}\label{eq:dot_product_2}
  \begin{split}
    \vec{v}_p=\frac{\vec{p}}{\bar{p}}\;,
    \quad\text{and}\quad
    \vec{v}_k=\frac{\vec{k}}{\bar{k}}\;.
  \end{split}
\end{equation}
The signed magnitudes of the momenta are defined as
\begin{equation}\label{eq:signed_velocities}
  \bar{p}=p_0\,\sqrt{1-\mu_p^2}\;,
  \quad\text{and}\quad
  \bar{k}=k_0\,\sqrt{1-\mu_k^2}\;,
\end{equation}
where
\begin{equation}\label{eq:signed_velocities_2}
  \mu_p^2=\frac{p^2}{p_0^2}\;,
  \quad\text{and}\quad
  \mu_k^2=\frac{k^2}{k_0^2}\;.
\end{equation}
The solutions of the square roots are taken such that the real part is maximized.
Evaluating Eq.~\eqref{eq:signed_velocities} in terms of $\mu_p^2$ and $\mu_k^2$
ensures that the velocities remain precise for (near) on-shell momenta.
The difference $1-\vec{v}_p\vec{v}_k$ leads to large cancellations
at small angles, which is one of the problems we aim to address.
A numerically stable computation can be performed with the
help of the identity~\cite{Kahan:2005aa,Kahan:2006aa}
\begin{equation}\label{eq:omct}
    2(1-\vec{v}_p\vec{v}_k)=
    (\vec{v}_p-\vec{v}_k)^2\;.
\end{equation}
We are left to evaluate the difference in the square bracket
of Eq.~\eqref{eq:dot_product} as accurately as possible.
This is achieved with the help of the following identity,
again making use of the numerical stability of $\mu_p^2$
and $\mu_k^2$,
\begin{equation}
  \begin{split}
    1-\frac{\bar{p}\,\bar{k}}{p_0k_0}=
    \frac{\mu_p^2+\mu_k^2-\mu_p^2\mu_k^2}{
    1+\sqrt{1-\mu_p^2\vphantom{\mu_k^2}}
    \sqrt{1-\mu_k^2\vphantom{\mu_p^2}}}\;.
  \end{split}
\end{equation}
In both vertices and propagators we often encounter terms of the form
$p_\pm=p_0\pm p_z$. In configurations with momenta collinear to the $z$-direction,
the evaluation of such expressions necessitates a reformulation in terms of large
and small components. This can be achieved using the following relation:
\begin{equation}\label{eq:stable_plus_minus}
  p_\pm=\left\{\begin{array}{cc}
  p_0\pm p_z &\text{if } \begin{array}{c}
  |p_0\pm p_z|^2>|p_0\mp p_z|^2\\[.25em]
  \text{or}\quad |p_z|^2<|p_\perp|^2\end{array}\\[1em]
  \displaystyle\frac{p^2+p_\perp^2}{p_0\mp p_z} &\text{else}
  \end{array}\right.\;.
\end{equation}
The condition $|p_z|^2<|p_\perp|^2$ is particularly relevant for the
polarization vectors that will be introduced in Eq.~\eqref{eq:polvecs_real}.
It is always fulfilled in the initial-state collinear regions.\\

In the deep infrared region of final-state collinear sectors of the phase space,
it is not possible to store external momentum configurations at sufficient
precision to allow for the reliable extraction of transverse or anti-parallel
components relevant to collinear splittings. These components, and the
related scalar invariants must be obtained from a phase-space generator
which computes them directly in terms of the integration variables.
Such generators can be constructed based on any collinear phase-space 
parametrization, for example~\cite{Frixione:1995ms,*Catani:1996vz,*Catani:2002hc}.
A convenient method for NNLO calculations was introduced in~\cite{Czakon:2010td,
  *Czakon:2011ve,*Caola:2017dug,*Asteriadis:2020gzh}. The supplemental material
of this article discusses how it is used in our numerical tests. 
A high-precision evaluation of the matrix element then requires
as input not only the momenta of the individual particles, but
also their difference to the collinear direction and
the invariants in the collinear sector.

\subsection{Wave functions of external states}
\label{sec:external_states}
The numerical stability of the external wave functions and their products with
vertex factors often plays a key role in numerical computations. It is particularly
important in initial- and final-state collinearly enhanced regions.
The spinors of external fermions are typically constructed using the Weyl
representation~\cite{Dixon:1996wi,*Dittmaier:1998nn}
\begin{equation}\label{eq:weyl_eigenspinors_wvdw}
  \begin{split}
    \chi_+(p)&=\left(\begin{array}{c}
      \sqrt{p_+}\\\sqrt{p_-}e^{i\phi_{p}}\end{array}\right)\;,\\
    \chi_-(p)&=\left(\begin{array}{c}
      \sqrt{p_-}e^{-i\phi_{p}}\\-\sqrt{p_+}\end{array}\right)\;.
  \end{split}
\end{equation}
These expressions are evaluated in a numerically stable fashion in a straightforward
manner, by using Eq.~\eqref{eq:stable_plus_minus} to compute the light-cone momenta
$p_+$ and $p_-$. The polarization vectors pose a slightly more difficult problem,
due to their gauge dependence. In the helicity formalism~\cite{Berends:1981rb,
  *DeCausmaecker:1981jtq,*Kleiss:1985yh,*Xu:1986xb,*Gunion:1985vca},
they contain a longitudinal component, which becomes problematic
if the angle between the gauge vector\footnote{In the helicity formalism,
the polarization vectors are given by $\varepsilon^\mu_{\pm}(p,q)=
\pm\langle q^\mp|\gamma^\mu|p^\mp\rangle/(\sqrt{2}\langle q^\mp|p^\pm\rangle)$,
with $q^\mu$ the gauge vector.} and the momentum is small.
This can be made explicit by assuming a gauge vector of $e_\pm^\mu=(1,0,0,\pm1)$,
in which case the polarization vector for helicity $\lambda$ is given by
\begin{equation}\label{eq:polvecs_explicit}
  \begin{split}
  \varepsilon^\mu_\lambda(p,e_\pm)=&\pm e^{-i\lambda\phi_p} 
  \bigg[\frac{p^\pm}{\sqrt{2p^+p^-}}\,e_\pm^\mu\\
    &\qquad\qquad+\frac{e^{\mp i\lambda\phi_p}}{\sqrt{2}}
    \bigg(e_x^\mu\pm i\lambda\, e_y^\mu\bigg)\bigg]\;,
  \end{split}
\end{equation}
with $e_x^\mu=(0,1,0,0)$, and $e_y^\mu=(0,0,1,0)$. One finds that the component
along $e_\pm^\mu$ diverges as $\sqrt{p^\pm/p^\mp}$. It is of course possible to
remove this artifact by using a gauge vector, $q$, with $(\vec{p}\vec{q})/(p_0q_0)<0$.
Nevertheless, we propose a different definition based on real-valued three-vectors,
due to its convenient numerical and analytical properties
\begin{equation}\label{eq:polvecs_real}
\begin{split}
  \varepsilon_1^\mu(p)=&\;\bigg(0,\frac{(\vec{n}\vec{p}\,)\vec{p}-(p_0^2-p^2)\vec{n}}{
    \sqrt{\vec{n}^2(p_0^2-p^2)-(\vec{n}\vec{p}\,)^2}\,\sqrt{p_0^2-p^2}}\bigg)\,,\\
  \varepsilon_2^\mu(p)=&\;\bigg(0,\frac{\vec{n}\times\vec{p}}{
    \sqrt{\vec{n}^2(p_0^2-p^2)-(\vec{n}\vec{p}\,)^2}}\,\bigg)\;,
\end{split}
\end{equation}
where $\vec{n}$ is $\vec{e}_{x}$, $\vec{e}_{y}$ or $\vec{e}_{z}$, depending on
whether the largest component of the momentum is in $x$-, $y$- or $z$-direction.
These real-valued polarizations can be used to define circular polarizations
as $\varepsilon_\pm^\mu(p)=(\varepsilon_1^\mu(p)\pm i\varepsilon_2^\mu(p))/\sqrt{2}$.

\subsection{Vertices and propagators}
\label{sec:vertices_propagators}
When the individual components of a polarization vector are large compared
to its virtuality, it becomes necessary to track the virtuality as an independent
variable in the calculation. Scalar products with other currents are then
evaluated using Eq.~\eqref{eq:dot_product}. Similarly, vertex factors involving
light-cone momentum fractions should be computed using Eq.~\eqref{eq:stable_plus_minus}.
Vertices involving fermions and fermion propagators are evaluated in the
same fashion.

The computation of off-shell gluon currents~\cite{BERENDS1988759} in the
collinear region of phase space requires special methods in order to avoid
numerical instabilities. Consider a pair of on-shell gluons with momenta $p_a^\mu$
and $p_b^\mu$ and polarization vectors $\varepsilon_a^\mu$ and $\varepsilon_b^\mu$. 
The space-time dependent part of the corresponding two-particle current
is generated by the triple-gluon vertex as follows:
\begin{equation}\label{eq:tgc}
  \begin{split}
    &\varepsilon_a^\mu\varepsilon_{b}^\nu\Gamma_{\mu\nu}^{\quad\rho}(p_a,p_b,-p_{ab})\\
    &\quad=(\varepsilon_a\varepsilon_b)(p_a-p_b)^\rho
    +(2p_b\varepsilon_a)\varepsilon_b^\rho-(2p_a\varepsilon_b)\varepsilon_a^\rho\;,
  \end{split}
\end{equation}
where $p_{ab}=p_a+p_b$. In the collinear limit $p_a\parallel p_b$,
the last two terms are nearly transverse, because they are proportional to the
one-particle currents. While formally transverse to $p_{ab}$, the first term,
in fact, has a large longitudinal component that can be made explicit in
a Sudakov decomposition~\cite{Sudakov:1954sw}
\begin{equation}\label{eq:sud_general}
  \begin{split}
    p_a^\mu=&\;z_a\tilde{p}_{ab}^\mu
    +\frac{{\rm k}_\perp^2+p_a^2}{z_a\,2\tilde{p}_{ab}n}\,
    n^\mu-k_\perp^\mu\;,\\
    p_b^\mu=&\;z_b\tilde{p}_{ab}^\mu
    +\frac{{\rm k}_\perp^2+p_b^2}{z_b\,2\tilde{p}_{ab}n}\,
    n^\mu+k_\perp^\mu\;.
  \end{split}
\end{equation}
Here, $n$ is an auxiliary, light-like vector that is linearly independent
of the gluon momenta, and we have defined $\tilde{p}_{ab}^\mu=p_{ab}^\mu
  -p_{ab}^2/(2p_{ab}n)\,n^\mu$, as well as ${\rm k}_\perp^2=-k_\perp^2$.
For practical applications, we choose $n=(\sqrt{\vec{p}_{ab}^{\,2}},-\vec{p}_{ab})$.
To improve the numerical stability of Eq.~\eqref{eq:tgc}, we perform
the calculation in a light-like axial gauge, which is particularly suitable
for collinear parton evolution~\cite{Amati:1978wx,*Amati:1978by,*Ellis:1978sf,
  *Ellis:1978ty,*Kalinowski:1980ju,*Kalinowski:1980wea,*Gunion:1985pg,*Catani:1999ss}.
In this gauge, the gluon polarization tensor reads
\begin{equation}\label{eq:axial_gauge}
  \begin{split}
    d^{\mu\nu}(p_{ab},n)=&\;-g^{\mu\nu}
    +\frac{p_{ab}^\mu n^\nu+n^\mu p_{ab}^\nu}{p_{ab}n}\;.
  \end{split}
\end{equation}
Multiplication by $(p_a-p_b)^\mu$ from the first term of Eq.~\eqref{eq:tgc}
yields the simple expression
\begin{equation}\label{eq:g_to_sqsq}
    d^\mu_{\;\rho}(p_{ab},n)\,(p_a-p_b)^\rho=2k_\perp^\mu
    +(z_a-z_b)\,\frac{p_{ab}^2}{p_{ab}n}\,n^\mu\;.
\end{equation}
Upon using the dedicated phase-space generator discussed earlier to parametrize
the collinear limit, the values of $k_\perp^\mu$ and $p_{ab}^2$ are known
to high precision.

Similar to the gluon currents, quark currents require a dedicated treatment
in the collinear limit. Consider the product of propagator numerator and
vertex factor for an on-shell quark, $a$, coupling to an on-shell gluon, $b$:
\begin{equation}
  \begin{split}
    (\slash\!\!\!p_a+\slash\!\!\!p_b)\slash\!\!\!\varepsilon_b\,u_a
    =(2p_a\varepsilon_b)u_a-i\sigma_{\mu\nu}p_b^\mu\varepsilon^\nu_b\,u_a\;.
  \end{split}
\end{equation}
The scalar interaction term can be evaluated using the methods described above.
The magnetic term creates numerical instabilities of a different type.
Using the Weyl representation of the $\gamma$-matrices, and the circular
polarizations based on Eq.~\eqref{eq:polvecs_real}, we obtain
\NiceMatrixOptions{cell-space-limits=1pt}
\begin{equation}\label{eq:magnetic}
  \begin{split}
    \frac{\sigma_{\mu\nu}p_b^\mu\varepsilon^\nu_{b,+}}{
      \sqrt{2}|p_{b\,\perp}|}=&\left(\;
      \begin{NiceArray}{c|cc}\Block{1-1}<\large>{0} & \Block{1-2}<\large>{0}\\\hline
        \Block{2-1}<\large>{0} & 1 & -p_b^+/p_{b\,\perp}\\
        & p_{b\,\perp}/p_b^+ & -1
        \end{NiceArray}\right)\;,\\
    \frac{\sigma_{\mu\nu}p_b^\mu\varepsilon^\nu_{b,-}}{
      \sqrt{2}|p_{b\,\perp}|}=&\left(\;
      \begin{NiceArray}{cc|c}
        -1 & -p_{b\,\perp}^*/p_b^+ & \Block{2-1}<\large>{0}\\
        p_b^+/p_{b\,\perp}^* & 1 & \\\hline
        \Block{1-2}<\large>{0} & & \Block{1-1}<\large>{0}
        \end{NiceArray}\right)\;.\\
  \end{split}
\end{equation}
In the exact collinear limit, $p_a=z_a p_{ab}$, $p_b=z_b p_{ab}$, 
the products of these matrices and the Dirac spinors derived from
Eq.~\eqref{eq:weyl_eigenspinors_wvdw} vanish. Numerically precise
results in final-state collinear regions can therefore only be
obtained by expressing $\sigma_{\mu\nu}p_b^\mu\varepsilon^\nu_b u_a$
in terms of Sudakov variables similar to Eq.~\eqref{eq:sud_general},
but separating the transverse momentum, $k_\perp$, into a component,
$k_\perp^+$, which is parallel to the $z$-axis, and a remainder
in the transverse plane.

\begin{figure*}[t]
    \centering
    \includegraphics[width=.475\textwidth]{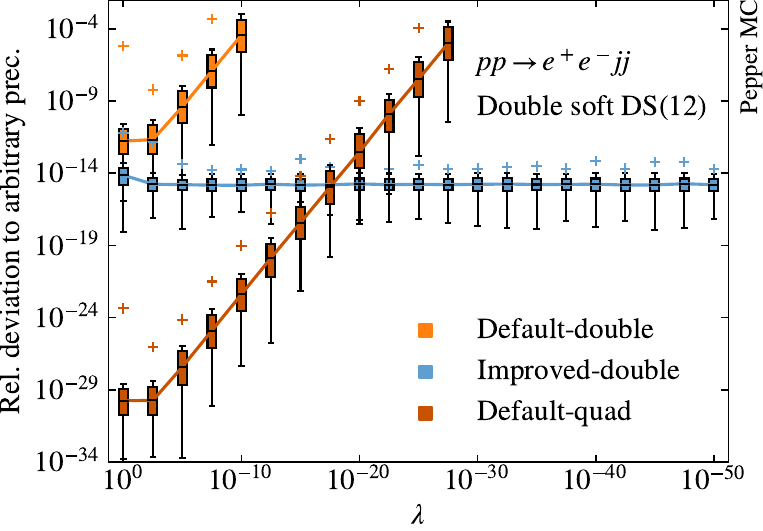}\hfill
    \includegraphics[width=.475\textwidth]{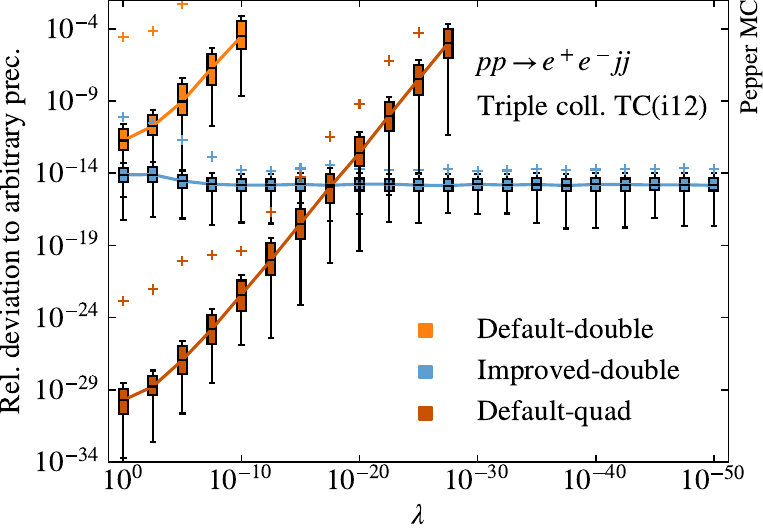}\\[3mm]
    \includegraphics[width=.475\textwidth]{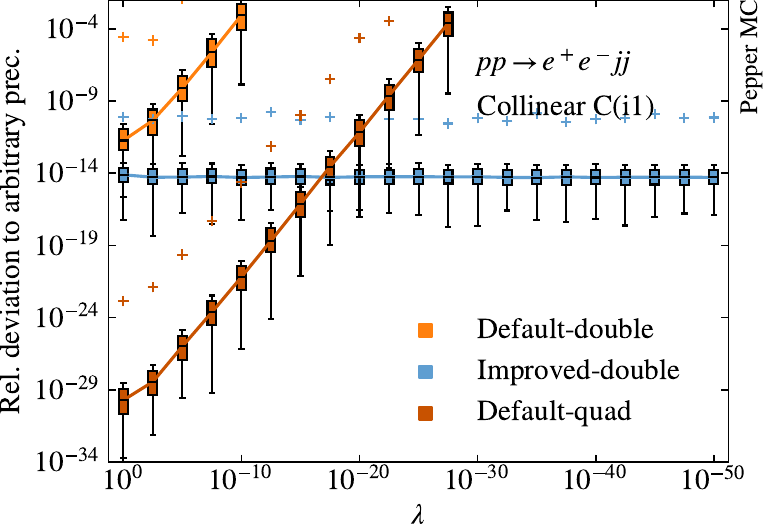}\hfill
    \includegraphics[width=.475\textwidth]{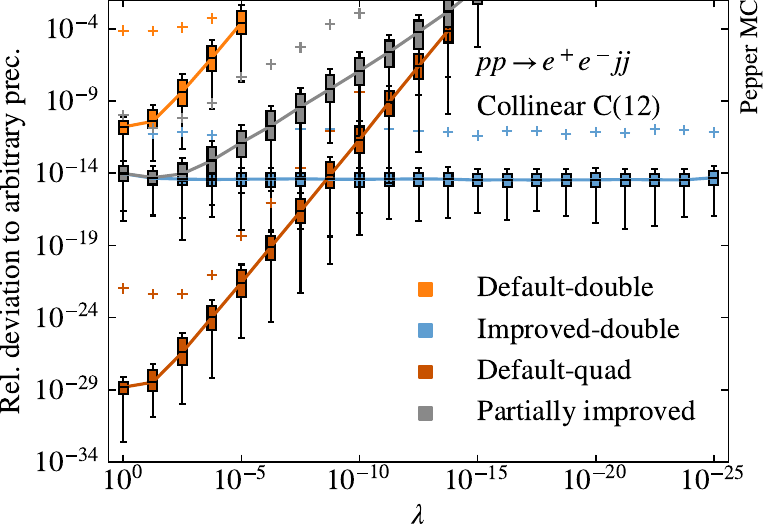}\\
    \caption{Relative precision of the matrix-element computation as a function
    of the scaling parameter, $\lambda$ (see supplemental material),
    in the double-soft (top left), 
    triple-collinear (top right), initial-state double-collinear (bottom left)
    and final-state double-collinear (bottom right) limit. We compare
    a naive double precision (orange), naive quadruple precision
    (red) and improved double precision (blue) implementation.
    The box plots are obtained from $10^4$ phase-space trajectories
    constructed as described in the supplemental material. The boxes show
    the quartiles of the distribution, the whiskers indicate $\nicefrac{3}{2}$
    of the interquartile range, and the crosses mark the maximum values.
    The grey distribution shows the improvements obtained in final-state
    collinear regions without a Sudakov decomposition of the spin-dependent
    part of interaction vertices.}
    \label{fig:limits}
\end{figure*}

\section{Example application: LHC physics}
\label{sec:tests}
We demonstrate the utility of our method using tree-level
matrix elements for Drell Yan lepton-pair production in association
with two jets. The reference value is obtained from a calculation in
arbitrary precision arithmetic with 256 significant digits, using the
\texttt{mp++} library~\cite{francesco_biscani_2023_10423945}.
Figure~\ref{fig:limits} compares the achieved precision relative to the
target from this implementation in the double-soft, triple-collinear
and double-collinear limits. We adopt the scaling procedure described
in the supplemental material of this article, which relies on the
phase-space parametrization of~\cite{Czakon:2010td,*Czakon:2011ve,
  *Caola:2017dug,*Asteriadis:2020gzh}.

\begin{table}[t]
    \centering
    \setlength\tabcolsep{2ex}
    \renewcommand{\arraystretch}{1.25}
    \begin{tabular}{p{8mm}|p{18mm}|p{18mm}|p{18mm}}
    $\tau_{\rm cut}$ & \multicolumn{3}{c}{
      $\delta\sigma_{\rm NNLO}({u\bar{u}}\to e^+e^-+X)$ [pb]} \\\cline{2-4}
    & \centering naive double & \multicolumn{2}{c}{improved double} \\\cline{3-4}
    & & \centering analytic ME & \multicolumn{1}{c}{BG recursion}  \\\hline
    $10^{-2}$ & 28.11(5) & 28.12(5) & 28.10(7) \\
    $10^{-3}$ & 27.4(3) & 27.0(1) & 27.0(1) \\
    $10^{-4}$ & 27.2(4) & 27.4(3) & 27.0(3) \\
    $10^{-5}$ & - & 27.2(9) & 27.1(8) \\
    $10^{-6}$ & - & 27.7(18) & 27.4(16) \\
    \end{tabular}
    \caption{Contribution of the $u\bar{u}$ initial state to the NNLO correction
    in $pp\to e^+e^-$ at $\sqrt{s}=13$~TeV. See the main text for details.}
    \label{tab:cross_sections}
\end{table}
In order to gauge the numerical stability of our algorithms
in a practical calculation, we perform a computation of the
$u\bar{u}$ contribution to the NNLO correction, $\delta\sigma_{\rm NNLO}$,
in Drell-Yan lepton pair production at the LHC at
$\sqrt{s}=13$~TeV. We use the PDF4LHC15+LUXqed NNLO PDF
set~\cite{Butterworth:2015oua,*Manohar:2017eqh}, and the corresponding
definition of the strong coupling. The electroweak parameters
are defined as $\sin^2\theta_W=0.2223$ and $\alpha=1/128.89$.
We set $m_Z=91.1876$~GeV and $\Gamma_Z=2.4952$~GeV.
The final-state electrons are required to fulfill
$66~{\rm GeV}\le m_{e^+e^-}\le 116~{\rm GeV}$.
Table~\ref{tab:cross_sections} shows the results for
$\delta\sigma_{\rm NNLO}$ in the $u\bar{u}$ channel,
computed using jettiness slicing~\cite{Boughezal:2016wmq,
  *Campbell:2019dru}. The results labeled ``BG recursion'' are obtained
with a combination of MCFM~\cite{Campbell:2015qma,*Campbell:2019dru}
and Pepper~\cite{Bothmann:2023gew}, all other
results stem from MCFM. The uncertainty on the results is driven
by the subtracted real emission corrections, for which
we employ the techniques described before, both in the real-emission
matrix element, and in all associated infrared counterterms.
The calculation reported in table~\ref{tab:cross_sections}
should be independent of the parameter $\tau_{\rm cut}$
in the limit $\tau_{\rm cut} \to 0$ and in practice
one must choose a value of $\tau_{\rm cut}$ that is deemed
to be in the asymptotic region. The naive double precision
implementation fails below $\tau_{\rm cut}\approx 10^{-4}$, 
leaving only two data points to support this assertion.
It is therefore not clear whether the result is correct. 
The improved implementation remains usable down to
$\tau_{\rm cut}\approx 10^{-6}$, and its results demonstrate
that the asymptotic region has indeed been reached.
The total additional computing time expended in our
unoptimized code is 5\% for MCFM, and 80\% for MCFM+Pepper.
We expect a reduction of this overhead in an optimized implementation.

\section{Conclusions}
\label{sec:conclusions}
This letter discussed a simple set of physics-driven algorithms
to stabilize the computation of tree-level matrix elements needed
for precision physics at colliders. The methods are applicable
to both analytically computed matrix elements and fully numerical
approaches. We have presented example numerical results
and demonstrated the utility of the technique for precision physics.
In view of the increased statistics and reduced experimental systematics
in the upcoming high-luminosity phase of the Large Hadron Collider,
the improvements discussed here will be of utmost importance for the
extraction of Standard Model parameters from experimental measurements.\\

This research was supported by the Fermi National Accelerator
Laboratory (Fermilab), a U.S.\ Department of Energy, Office of
Science, HEP User Facility managed by Fermi Research Alliance,
LLC (FRA), acting under Contract No. DE--AC02--07CH11359.
The work of J.M.C. and S.H. was supported by the U.S. Department of Energy,
Office of Science, Office of Advanced Scientific Computing Research,
Scientific Discovery through Advanced Computing (SciDAC-5) program,
grant “NeuCol”. E.B. and M.K. acknowledge support from BMBF 
(contract 05H21MGCAB). Their research is funded by the 
Deutsche Forschungsgemeinschaft (DFG) – 456104544; 510810461.
We used resources of the National Energy Research 
Scientific Computing Center (NERSC), a U.S.\ Department of Energy
Office of Science User Facility using NERSC awards HEP-ERCAP0023824
and HEP-ERCAP0028985.

\onecolumngrid
\bibliography{main}

\clearpage
\widetext
\begin{center}
\textbf{\large Supplemental Material: Algorithms for numerically stable scattering amplitudes}
\end{center}
\setcounter{equation}{0}
\setcounter{figure}{0}
\setcounter{table}{0}
\setcounter{page}{1}
\makeatletter
\renewcommand{\theequation}{S\arabic{equation}}
\renewcommand{\thefigure}{S\arabic{figure}}

\subsection{Scaling procedure}
To generate the phase-space trajectories approaching the infrared limits relevant to NNLO
calculations, we use the phase-space parametrization of~\cite{Czakon:2010td,*Czakon:2011ve,
  *Caola:2017dug,*Asteriadis:2020gzh}. The construction starts with a hard momentum, $p_i$,
which defines the collinear direction. The momenta of up to two emissions, $p_1$ and $p_2$,
can be written as
\begin{equation}
  \begin{split}
    p_1^\mu=&\;t^\mu+\cos\theta_{i1} e_i^\mu+\sin\theta_{i1} b^\mu\;,\\
    p_2^\mu=&\;t^\mu+\cos\theta_{i2} e_i^\mu+\sin\theta_{i2} (\cos\phi\, b^\mu + \sin\phi\, a^\mu)\;,
  \end{split}
\end{equation}
where $t^\mu=(1,\vec{0})$, $e_i^\mu=(0,\vec{p}_i/|\vec{p}_i|)$, $b^\mu$ is transverse to $t^\mu$ and $e_i^\mu$,
and $a^\mu$ is transverse to $t^\mu$, $e_i^\mu$ and $b^\mu$. The construction of the transverse vectors
is achieved as follows
\begin{equation}
  \begin{split}
    b^\mu=&\;\frac{\tilde{b}^\mu}{\sqrt{-\tilde{b}^2}}\;,
    \qquad\text{where}\qquad
    \tilde{b}^\mu=\sum_{k=1,3}\varepsilon^\mu_{\;\;\nu\rho k}\,e_i^\nu t^\rho\;,
    \qquad\text{and}\qquad
    a^\mu=\;\varepsilon^\mu_{\;\;\nu\rho\sigma}\,b^\nu e_i^\rho t^\sigma\;.
  \end{split}
\end{equation}
One defines the angular variables
\begin{equation}
    \begin{split}
        \eta_{1/2}=\eta_{i1/i2}=\frac{1-\cos\theta_{i1/i2}}{2}
        \qquad\text{and}\qquad
        \eta_{12} = \frac{1-\cos\theta_{12}}{2}\;.
    \end{split}
\end{equation}
In terms of the azimuthal angle $\phi_{12}$, the variable $\eta_{12}$ is given by
\begin{equation}
    \eta_{12}=\eta_1+\eta_2-2\eta_1\eta_2-2\cos\phi_{12}\sqrt{\eta_1(1-\eta_1)\eta_2(1-\eta_2)}\;.
\end{equation}
A naive parametrization of the azimuthal integral would be given by $\cos\phi_{12}=1-2\xi$.
However, in order to disentangle the singularities at small $\eta_{1/2}$ and small $\eta_{12}$,
one uses instead~\cite{Czakon:2010td}
\begin{equation}
    \zeta=\frac{1-\xi}{1+2\xi X}\;,
    \qquad\text{where}\qquad
    X=\left[\,\frac{\eta_1+\eta_2-2\eta_1\eta_2}{2\sqrt{\eta_1(1-\eta_1)\eta_2(1-\eta_2)}}-1\,\right]^{-1}\;,
\end{equation}
such that
\begin{equation}
    \eta_{12}=\frac{(\eta_1-\eta_2)^2}{\eta_1+\eta_2-2\eta_1\eta_2-2(1-2\zeta)\sqrt{\eta_1(1-\eta_1)\eta_2(1-\eta_2)}}\;,\qquad
    \sin^2\phi_{12}=4\zeta(1-\zeta)\,\frac{\eta_{12}^2}{(\eta_1-\eta_2)^2}\;.
\end{equation}
Starting with a set of phase-space points produced by a diagram-based phase-space
generator~\cite{Sherpa:2019gpd}, and a set of indices corresponding to $i$, $1$ and $2$,
we order the indices such that the momenta satisfy $E_2<E_1$. We then determine the angles
$\eta_{1/2}$, $\xi$, and the transverse vectors $a^\mu$ and $b^\mu$. We map the energies
and angles to the parameters $x_{1/2}$ and $x_{3/4}$, using $E_{1}=x_1 E_{{\rm max}}$,
$E_{2}=x_1 x_2 E_{{\rm max}}$, and one of the following angular
parametrizations~\cite{Caola:2017dug,Asteriadis:2020gzh}:
\begin{equation}\label{eq:nscs_params}
  \begin{split}
  \text{Sector (a)}:&& \eta_{1} =&\; x_3\,, &\eta_{2} =&\; \frac{x_3x_4}{2}\;,\qquad&
  \text{Sector (b)}:&& \eta_{1} =&\; x_3\,, &\eta_{2} =&\; x_3\Big(1-\frac{x_4}{2}\Big)\;,\\
  \text{Sector (c)}:&& \eta_{2} =&\; x_3\,, &\eta_{1} =&\; \frac{x_3x_4}{2}\;,\qquad&
  \text{Sector (d)}:&& \eta_{2} =&\; x_3\,, &\eta_{1} =&\; x_3\Big(1-\frac{x_4}{2}\Big)\;.
  \end{split}
\end{equation}
There is exactly one such mapping for each phase-space point.
We can then scale the original point into the soft ($S$), double soft ($DS$),
double collinear ($C$), and triple collinear ($TC$) limits by redefining the variables
$x_1$ through $x_4$, and reconstructing the original point using the new variables. The
possible options for rescaling are listed in Tab.~\ref{tab:scaling}. They can be combined
in order to test strongly ordered regions. For example, the combination $C(12)DS(12)$
corresponds to the collinear limit of a double-soft $(12)$ particle pair.
Note that in the double-collinear limit, $x_4\to 0$, the angular variable $\eta_{12}$
in sectors (b) and (d) is proportional to $x_4^2$, such that the $12$-collinear limit
is approached twice as fast as the $i1$ and $i2$ collinear limits in sectors (a) and (c).
As intended in the original publications~\cite{Czakon:2010td,*Czakon:2011ve,
  *Caola:2017dug,*Asteriadis:2020gzh}, the above algorithm can also be used directly
as a phase-space generator for collinear configurations.
\begin{table}
\setlength{\tabcolsep}{6pt}
\renewcommand{\arraystretch}{1.25}
\begin{tabular}{p{.1\textwidth}|p{.125\textwidth}|p{.075\textwidth}c
                p{.1\textwidth}|p{.125\textwidth}|p{.075\textwidth}}
Limit & Scaling & Sector & \hspace*{3mm} &
Limit & Scaling & Sector \\\cline{1-3}\cline{5-7}
$S(2)$ & $x_2\to\lambda$ & any &&
$DS(12)$ & $x_1\to\lambda$ & any \\
$C(i1)$ & $x_4\to\lambda$ & (a) &&
$TC(i12)$ & $x_3\to\lambda$ & any \\
$C(i2)$ & $x_4\to\lambda$ & (c) &&
$C(12)$ & $x_4\to\lambda$ & (b) or (d) \\
\end{tabular}
\caption{Scaling of the phase-space parameters in Eq.~\eqref{eq:nscs_params}
in the soft ($S$), double soft ($DS$), double collinear ($C$),
and triple collinear ($TC$) limits. The above limits can be combined into
strongly ordered limits by scaling two parameters.\label{tab:scaling}}
\end{table}

\subsection{Infrared subtracted real-emission corrections}
In this subsection we discuss the numerical precision of infrared subtracted
real-emission QCD corrections at NLO that can be achieved with our new algorithms.
We make use of the Catani-Seymour dipole method~\cite{Catani:1996vz}.
When evaluating the corresponding dipole subtraction terms, it is important
to use the stable methods for computation of all potentially small dot products
in the dipole insertion operators. Moreover, terms such as $1-z$ should be evaluated
as a whole in terms of dot products whenever $z$ can be large.
These modifications are straightforward. The only nontrivial adaptation concerns
the four-vectors which are used to evaluate the transverse momentum dependent parts
of the gluon splitting functions. They are defined for final-final (FF),
final-initial (FI), initial-final (IF) and initial-initial (II) dipoles in
Eqs.~(5.8), (5.41), (5.67) and~(5.147) of Ref.~\cite{Catani:1996vz}, respectively:
\begin{equation}\label{eq:dippol}
  \begin{split}
    k^\mu_{\perp,\rm FF/FI}=&\;\frac{\tilde{z}_i p_i^\mu-\tilde{z}_j p_j^\mu}{
      \sqrt{\tilde{z}_i\tilde{z}_j\, 2p_ip_j}}\;,
    \qquad\text{and}\qquad &
    k^\mu_{\perp,\rm IF/II}=&\;\sqrt{\frac{p_ip_a\,p_ap_k}{2p_ip_k}}
    \bigg(\frac{p_i^\mu}{p_ip_a}-\frac{p_k^\mu}{p_kp_a}\bigg)\;.
  \end{split}
\end{equation}
These vectors are formally transverse to the emitter momenta,
$\tilde{p}_{ij}$ and $\tilde{p}_a$, and normalized to $k_\perp^2=-1$.
However, a Sudakov decomposition reveals that they contain
a non-vanishing component in the direction of the emitter, and that
this component diverges in the collinear limit, due to the normalization.
Current conservation in the on-shell matrix elements eliminates
terms proportional to $\tilde{p}_{ij}^\mu$ and $\tilde{p}_a^\mu$
in the insertion operators, however, numerically this
is achieved through large cancellations in scalar products
of near collinear momenta. We use an axial gauge to remove the
longitudinal components in $k_\perp^\mu$ explicitly, which corresponds
to subtracting $(\tilde{z}_i-\tilde{z}_j)\,\tilde{p}_{ij}^\mu/
 \sqrt{\tilde{z}_i\tilde{z}_j\,2p_ip_j}$ in final-state dipoles,
and $\sqrt{p_ap_k/(2p_ip_k\,p_ip_a)}\,(1-x)\,p_a^\mu$ in initial-state dipoles.
This method stabilizes the calculation at $\lambda\lesssim 10^{-9}$.

\begin{figure}
    \centering
    \includegraphics[width=\textwidth]{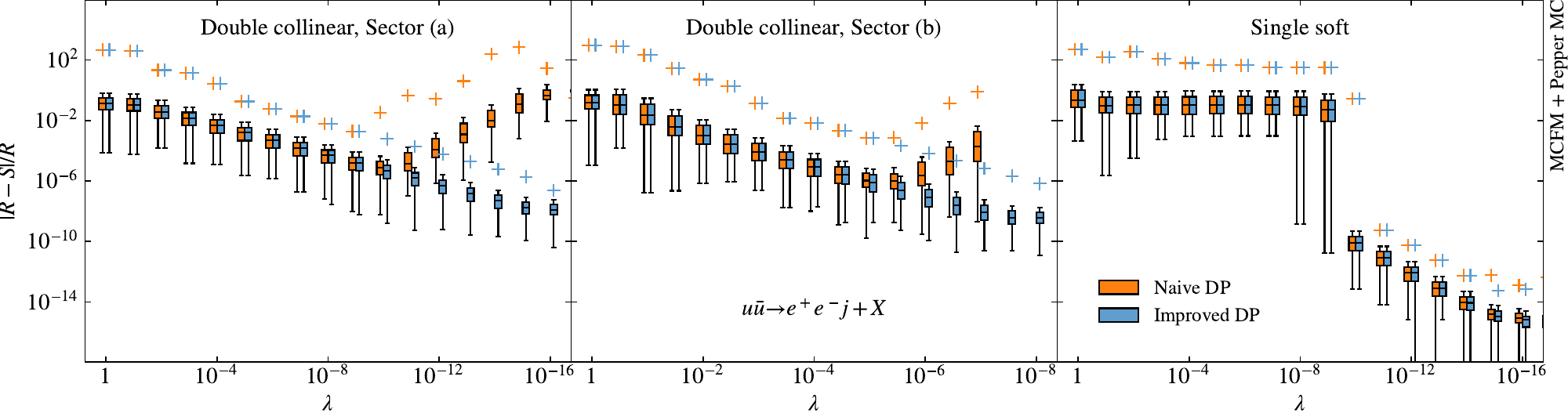}
    \caption{Absolute value of the relative difference between the infrared subtraction
    terms and the real-emission matrix elements as a function of the scaling parameters,
    $\lambda$. We compare a naive double precision implementation (Naive DP) and the
    improved algorithm (Improved DP). The boxes show the median and quartiles of the
    distribution in $10^4$ phase-space trajectories,
    the whiskers indicate $\nicefrac{3}{2}$
    of the interquartile range
    and the crosses show the maximum values.
    For details see the main text.}
    \label{fig:rsub_cancellation}
\end{figure}

Figure~\ref{fig:rsub_cancellation} shows tests of the cancellation between
the dipole subtraction terms and the real-emission correction in the
$u\bar{u}$ initial state contribution to the NNLO double-real
correction above $\tau_{\rm cut}=10^{-6}$ in the jettiness slicing approach.
The plots are obtained from $10^4$ phase-space points, which are scaled
according to the procedure described above.
The left panel shows the behavior in the initial-state double collinear limit,
the middle panel in the final-state double collinear limit, and the right panel
in the single soft limit. We find that our improved numerical algorithms allow
to probe the infrared region in much more detail than previously possible.
The ledge in the right plot is due to dipole subtraction terms corresponding
to the soft limit on gluon 1 when the momentum of gluon 2 is scaled to zero.
These terms exhibit a soft scaling, leading to a constant offset
in Fig.~\ref{fig:rsub_cancellation} for $\tau(\{\tilde{p}\})>\tau_{\rm cut}$.
\end{document}